# Improved Reduced Models for Single-Pass and Reflective Semiconductor Optical Amplifiers

Seán P. Ó Dúill and Liam P. Barry

*Abstract—* We present highly accurate and easy to implement, improved lumped semiconductor optical amplifier (SOA) models for both single-pass and reflective semiconductor optical amplifiers (RSOA). The key feature of the model is the inclusion of the internal losses and we show that a few subdivisions are required to achieve an accuracy of 0.12 dB. For the case of RSOAs, we generalize a recently published model to account for the internal losses that are vital to replicate observed RSOA behavior. The results of the improved reduced RSOA model show large overlap when compared to a full bidirectional travelling wave model over a 40 dB dynamic range of input powers and a 20 dB dynamic range of reflectivity values. The models would be useful for the rapid system simulation of signals in communication systems, i.e. passive optical networks that employ RSOAs, signal processing using SOAs and for implementing digital back propagation to undo amplifier induced signal distortions.

## I. INTRODUCTION

MODELLING SEMICONDUCTOR optical amplifiers (SOA) has been a topic for over two decades [1]-[4]. Very recently, models of reflective SOAs (RSOA) have emerged, [3][4], mainly driven by RSOA exploitation within passive optical networks (PON) [5],[6]. Reduced (or lumped) SOA models, [1]-[3], allow for solving the gain and refractive index dynamics without having to solve computationally intensive propagation equations as was done in [4]. In all the reduced SOA models presented, the inclusion of internal scattering losses in the lumped SOA models have proven to be elusive due to the fact that no analytical solution arises when the internal scattering losses are non-zero [1],[7].

In this letter we propose an improved reduced model for both SOAs and RSOAs that approximates the inclusion of the internal scattering losses. The assumption is based on regarding the SOA's gain coefficient to be constant over a certain length of SOA. This assumption is certainly valid for: short SOA sections; when the optical power is much less than the SOA saturation power and under strong saturation conditions when the gain is depleted to the extent such that there are no large longitudinal variations in the gain coefficient. For single pass SOAs over a 50 dB dynamic range of input powers: we show that by even considering a single section; that the maximum discrepancy of 1 dB was found when calculating output power by considering a single calculation step over the entire SOA as opposed to splitting the SOA up into 40 subsections. The method obviates the need for a fine-grained SOA model, allowing for rapid and accurate system calculations of signal propagation through SOAs.

The improved reduced model for SOAs is extended to RSOAs and builds upon the simpler of two recently published reduced RSOA models [3], allowing for the inclusion of the internal scattering losses. The results from the improved reduced model are compared with the full travelling wave model (TWM) [4], showing excellent agreement with discrepancies < 1 dB over a 40 dB dynamic range of input powers combined with a 20 dB dynamic range of reflectivity values. In addition, we also show how the losses are incorporated in accounting for the intraband contributions to the nonlinear gain. The inclusion of these effects enables the simulation of all-optical signal processing using four-wave mixing (FWM).

## II. Improved Reduced SOA Model

We begin the analysis by transcribing the SOA propagation and gain dynamics equations from [1].

$$\frac{d|E(z,t)|^2}{dz} = \left(g(z,t) - \alpha_{loss}\right)|E(z,t)|^2 \quad (1)$$

$$\frac{d\phi(z,t)}{dz} = -\tfrac{1}{2}\alpha_H g(z,t) \quad (2)$$

$$\frac{dg(z,t)}{dt} = \frac{g_0 - g(z,t)}{\tau_S} - \frac{g(z,t)|E(z,t)|^2}{\tau_S P_{sat}} \quad (3)$$

Eqs.-(1) and (2) describe the amplification and phase shift accumulation of the optical field $E(z,t)$ along the SOA with $z$ and $t$ being the spatial and temporal variables; the optical power is given by $|E(z,t)|^2$. $\alpha_{loss}$ describes the internal scattering losses. $g$ is the gain coefficient whose dynamics are described by Eq-(3); the second term on the right hand side of which describes gain depletion due to stimulated emission while the first term describes gain recovery back to the unsaturated value $g_0$. The gain recovery time, $\tau_S$, is the carrier lifetime. $P_{sat}$ is the saturation power. The gain-phase coupling is determined by $\alpha_H$. The reduced models rely on integrating the gain over a length, $L$, of the SOA. Equating $h$ as the spatially-integrated SOA gain coefficient over $L$

This work is supported by Science Foundation Ireland PI program under grant No. (09/IN 1/12653).

S. Ó Dúill, and L. P. Barry are with the Rince Institute, School of Electronic Engineering, Dublin City University, Dublin 9, Ireland (email: sean.oduill@dcu.ie, liam.barry@dcu.ie ).

TABLE I
SOA PARAMETERS USED IN INITIAL VERIFICATION

| Symbol | Definition | Value |
|---|---|---|
| $h_0$ | Unsaturated gain parameter | 10.5 |
| $P_{sat}$ | Saturation power | 10 mW |
| $\tau_S$ | Carrier lifetime | 100 ps |
| $\alpha_H$ | Gain-phase coupling parameter | 5 |
| $\alpha_{loss}L$ | Internal losses | 4 |

$$h(t) = \int_0^L g(z,t)dz \equiv g_{av}(t)L, \quad (4)$$

we define $g_{av}(t)$ as the spatially-averaged gain coefficient. The assumption is valid as long as the spatial profile of the gain coefficient is constant. In principle, unidirectional signal amplification along the SOA causes the gain coefficient to monotonically decrease along the length of the SOA, thus requiring for the SOA to be broken up into many sections in order to capture the correct gain profile. Assuming a constant gain coefficient allows us to write an approximate analytical expression for the integral of the second term on the right hand side of Eq.-(3). The input optical field to the SOA is given as $E_{in}(t) = E(0,t)$; using (4), Eqs.-(1) and (3) can be re-written as:

$$|E(z,t)|^2 \approx |E_{in}(t)|^2 \exp\{(g_{av}(t) - \alpha_{loss})z\} \quad (5)$$

$$\frac{dh(t)}{dt} \approx \frac{h_0 - h(t)}{\tau_S} - \frac{g_{av}(t)|E_{in}(t)|^2 \int_0^L \exp\{(g_{av}(t) - \alpha_{loss})z\}dz}{\tau_S P_{sat}} \quad (6)$$

The integral in (6) can be performed by inserting the result for $|E(z,t)|^2$ from (5) and substituting $h$ for $g_{av}$ using (4) to give:

$$\frac{dh(t)}{dt} \approx \frac{h_0 - h(t)}{\tau_S} - \frac{h(t)}{h(t) - \alpha_{loss}L} \frac{\left[\exp\{h(t) - \alpha_{loss}L\} - 1\right]|E_{in}(t)|^2}{\tau_S P_{sat}} \quad (7)$$

An expression for the total phase change is written as

$$\phi_{tot}(t) = -\tfrac{1}{2}\alpha_H h(t) \quad (8)$$

The output optical field is simply expressed as

$$E_{out}(t) = E_{in}(t)\exp\{\tfrac{1}{2}(-\alpha_{loss}L + (1-j\alpha_H)h(t))\} \quad (9)$$

It should be noted that Eqs.-(7) and (9) reduce to the equations in [1] if $\alpha_{loss} = 0$; because the true analytical solution for $h(t)$ holds irrespective of the spatial gain profile $g(z,t)$ [1].

To verify Eqs.-(7) and (9), and to highlight the improvement of the current approach, we subdivide the SOA into separate sections and note the number of required subsections before the output power reaches a consistent value. The scenario is depicted in Fig. 1 with $K$ being the number of considered subsections. This is performed for continuous wave (CW) signals whose input power ranges from -40 to 10 dBm, with $K$ varying from 1 to 40. The results are shown in Fig. 2 using the SOA parameters given in Table I; the net unsaturated SOA gain is ~28 dB. We define the discrepancy between the output power calculation by considering $K$ subdivisions and 40 subdivisions as:

$$D(K) = 10\log_{10}\left(\frac{P_{out}^{(K)}}{P_{out}^{(40)}}\right). \quad (9)$$

As predicted for low input powers, there is no discrepancy

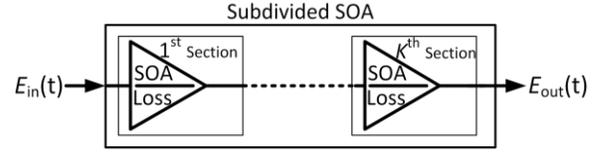

Fig. 1. Simulation setup employed to show the splitting of the SOA into $K$ subsection SOAs. The optical field output for the $(k-1)^{th}$ section becomes the input to the $k^{th}$ section.

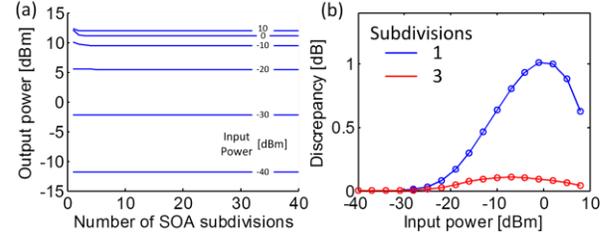

Fig. 2 (a) Plot of the calculated SOA output power vs. the number of SOA subdivisions used in the calculation; the input power is used as parameter. (b) The discrepancy when calculating the SOA output power using 1 and 3 subdivisions compared with the calculation by splitting the SOA up into 40 subsections is shown. A maximum 1 dB discrepancy in the output power is found by employing a single subdivision as opposed to 40 subdivisions. The discrepancy decreases dramatically below 0.12 dB by considering 3 subdivisions.

between the output power calculation because the SOA gain profile is constant because the power in the SOA is much less than $P_{sat}$. Though when the input power increases, the gain profile no longer remains flat thus requiring more subdivisions to get an accurate value for the output power. Though as is clear from Fig. 2, a maximum discrepancy of just 1 dB is found over the entire input power range up to 10 dBm by considering a single subdivision, this is an acceptable error in most circumstances. For the cases when greater accuracy is required, the discrepancy could be reduced below 0.12 dB by only considering 3 subdivisions, as is evident from Fig. 2.

The reduced model in [2] also accounted for the intraband contributions to the nonlinear SOA gain [7]. FWM is the only 3$^{rd}$ order nonlinear process that is transparent to modulation format and has been used to process a variety of signals with amplitude and/or phase encoding [2],[8],[9]. We will now show how the intraband effects can be included in the improved model. Starting with the rate equation describing the dynamics of carrier heating (CH) [2][7]:

$$\frac{d\Delta g_{ch}(t)}{dt} = -\frac{\Delta g_{ch}(t)}{\tau_{ch}} - \frac{g(t)|E(z,t)|^2}{P_{sat\_ch}\tau_{ch}} \quad (10)$$

where $\Delta g_{ch}$ is the gain change due to CH, $\tau_{ch}$ is the associated time constant with carrier-phonon collisions and is ~500 fs. $g$ is the optical gain defined in Eq.-(1) and $P_{sat\_ch}$ is the saturation powers associated with CH. Using the technique outlined in (4)-(7) and invoking the adiabatic condition that changes in $|E|^2$ occur over timescales longer than $\tau_{ch}$ i.e. $d\Delta g_{ch}/dt = 0$, then the spatially integrated version of (10) yields the contribution to the gain of:

$$\Delta h_{ch}(t) \approx -\frac{h(t)}{h(t) - \alpha_{loss}L} \frac{\left[\exp\{h(t) - \alpha_{loss}L\} - 1\right]|E_{in}(t)|^2}{P_{sat\_ch}} \quad (11)$$

TABLE II
SOA PARAMETERS USED IN FWM CALCULATIONS

| Symbol | Definition | Value |
| --- | --- | --- |
| $h_0$ | Unsaturated gain parameter | 11 |
| $P_{sat}$ | Saturation power | 40 mW |
| $\tau_S$ | Carrier lifetime | 60 ps |
| $\alpha_H$ | Gain-phase coupling parameter | 6 |
| $\alpha_{loss}L$ | Internal losses | 2 |
| $P_{sat\_ch}$ | Saturation power for carrier heating | 300 mW |
| $\alpha_{ch}$ | Carrier heating gain-phase coupling | 3 |

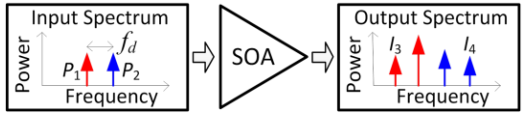

Fig. 1 Scenario for FWM. Two pumps $P_1$ and $P_2$ are input into the SOA. Idlers $I_3$ and $I_4$ are created due to the nonlinear interaction between the pumps.

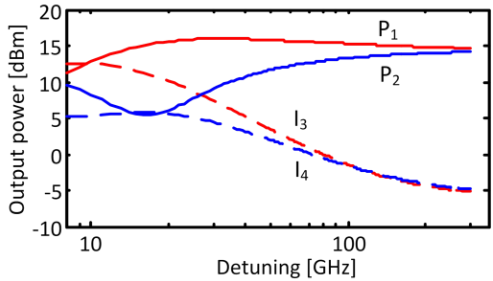

Fig. 2 Calculated output power of the pumps and idlers using the improved SOA model.

A similar expression could be written for spectral hole burning (SHB) [2][7]. For FWM, solving (10) in the adiabatic limit restricts signal-pump detunings to be less than 300 GHz, though this allows us to include the intraband effects without having to excessively oversample the input field to calculate $\Delta h_{ch}(t)$ using the spatially integrated form of (10). The output optical field is given by:

$$E_{out}(t) \approx E_{in}(t)\exp\{\tfrac{1}{2}(-\alpha_{loss}L + (1-j\alpha_H)h(t) + (1-j\alpha_{ch})\Delta h_{ch}(t) + \Delta h_{shb}(t))\} \quad (12)$$

With $\alpha_{ch}$ describing the refractive index dynamics associated with CH. The contribution arising from SHB is given by $\Delta h_{shb}$. We replicate the carefully obtained experimental results presented in [8] using the SOA parameters given in Table II. The situation is shown in Fig. 1 with two equal power pumps $P_1$ and $P_2$ =100 µW at the SOA input. The two pumps interact in the SOA creating two SOA idlers, $I_3$ and $I_4$ via FWM. The SOA input field is given as:

$$E_{in}(t) = \sqrt{P_1} + \sqrt{P_2}\exp(j2\pi f_d t) \quad (12)$$

The detuning, $f_d$, is varied from 8 to 300 GHz. The power of the pumps and idlers are extracted from the calculated output spectrum. The results are shown in Fig. 2 and agree quite well with the experimental and travelling-wave simulation results in [8]. The output power for both pumps show quite strong cross-gain modulation for detunings <100 GHz, with $P_1$ emerging considerably stronger. The two idlers also behave as was measured [8]. For detunings < 30 GHz, both idlers become commensurate with the pumps indicating that the device exhibits fast carrier dynamics.

TABLE III
RSOA PARAMETERS

| Symbol | Definition | Value |
| --- | --- | --- |
| $h_0$ | Unsaturated gain parameter | 10.5 |
| $P_{sat}$ | Saturation power | 10 mW |
| $\tau_S$ | Carrier lifetime | 100 ps |
| $\alpha_H$ | Gain-phase coupling parameter | 5 |
| $\alpha_{loss}L$ | Internal losses | 4.5 |
| $R$ | Reflectivity | 0.01 – 1 |
|  | Fiber-RSOA coupling loss | 3 dB |

### III. REFLECTIVE SOAs

RSOAs are finding applications as low-cost upstream transmitters within PONs [5]-[6],[10]; therefore targeting accurate, yet simple, models that describe their behavior is a laudable goal for system analysis simulations. In general, the output power characteristics of reflective amplifiers differ from single pass amplifiers in that a maximum output power is reached at a certain input power [11], and this was shown experimentally and numerically for RSOAs [10],[12]. Simplified travelling wave models for RSOAs have appeared [3],[4], with the model in [3] introducing a lumped RSOA model that shows nice agreement with the full travelling wave model in [4]. We show how the losses could be implemented in a reduced RSOA model using the techniques outlined in section II and we assume sub-unity reflectivity values for the reflective facet, as they are preferred to avoid lasing [13]. A depiction of an RSOA is shown in Fig. 3. The fields at the input and reflective facets are defined, and the superscript [+,-] defines the shown directions of propagation. A typical plot of the distributed nature of the localized RSOA gain is shown and the gain (carrier density) is greatest in the center of the device because the gain saturation is strongest at the input and reflective facets because the counter-propagating fields are strongest at the facets [12]. To model the RSOA gain dynamics we invoke the criterion for the reduced RSOA model in [3] that the intensity of the input signal does not vary over timescales of signal time-of-flight in RSOAs (~10 ps); such criterion is easily met by signals in WDM-PON scenarios with baudrates ≤ 10 Gbaud. Noting that the gain is depleted by two counter propagating waves originating from the two facets and including the internal losses, the integrated gain

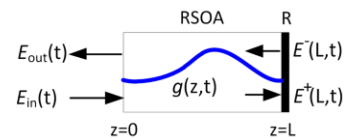

Fig. 3 Schematic of amplification within an RSOA. The input field $E_{in}$ travels along the RSOA. At $z = L$ the field is partially reflected back in the opposite direction and is amplified until the field re-emerges, $E_{out}$, at the z=0 facet. The maximum gain occurs near the middle of the RSOA [12].

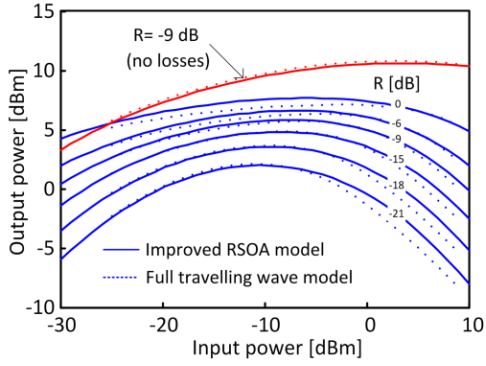

Fig. 4 Calculated RSOA input-output power transfer characteristics using the improved model and the full TWM using 30 subsections [4]. The red curves highlight show that internal losses are needed to accurately simulate RSOA.

dynamics are described by:

$$\frac{dh(t)}{dt} \approx \frac{h_0 - h(t)}{\tau_S} - \frac{h(t)}{h(t) - \alpha_{loss} L} \times \frac{\left[\exp\{h(t) - \alpha_{loss} L\} - 1\right]\left[|E_{in}(t)|^2 + |E_L^-(t)|^2\right]}{\tau_S P_{sat}} \quad (13)$$

To relate $E^-(L,t)$ to $E_{in}(t)$, we note that $E^+(L,t)$ is $E_{in}(L,t)$ after traversing the RSOA to $z = L$. The boundary condition at the reflective facet is $|E_L^-(t)|^2 = R|E_L^+(t)|^2$; thus giving an expression for $E^-(L,t)$ in terms of the input $E_{in}(t)$:

$$|E_L^-(t)|^2 = |E_{in}^+(t)|^2 R \exp\{h(t) - \alpha_{loss} L\} \quad (14)$$

Inserting (14) into (13) and rearranging gives the RSOA gain dynamics in response an arbitrary input optical field:

$$\frac{dh(t)}{dt} \approx \frac{h_0 - h(t)}{\tau_S} - \frac{h(t)}{h(t) - \alpha_{loss} L} \times \frac{\left[\exp\{h(t) - \alpha_{loss} L\} - 1\right]\left[1 + R \exp\{h(t) - \alpha_{loss} L\}\right]|E_{in}(t)|^2}{\tau_S P_{sat}} \quad (15)$$

Eq.-(15) reduces to the model in [3] when $R = 1$ and $\alpha_{loss} = 0$. The output optical field is simply as the amplification of the input field on the rightwards journey, reflected by the reflective facet (14) and amplified on the leftwards journey:

$$E_{out}(t) \approx E_{in}(t)\sqrt{R} \exp\{(-\alpha_{loss} L + (1 - j\alpha_H)h(t))\} \quad (16)$$

Using Eqs.-(15) and (16) we now present results of RSOA gain saturation for CW input signals varying from -30 to +10 dBm for with reflectivity $R$ ($-21 \leq R \leq 0$ dB) set to mimic the experimentally obtained curves in [10]. The calculated results in Fig. 6 were obtained using the RSOA parameters given in Table III for both the improved model and for the full TWM with the RSOA split into 30 sections and internal losses included with the TWM described in [3][4]. The RSOA length was taken to be 800 μm when implementing the TWM. There is excellent agreement between the obtained input-output power transfer characteristics from both models and both models replicate the findings of RSOA with differing values of reflectivity [10]. The remarkable thing is that there is considerable overlap between the curves with a maximum discrepancy of 1 dB despite the fact that the reduced RSOA model neither considers spatially-resolved counter-propagating fields nor spatially resolves the gain profile. All discrepancies are within ±1 dB over the entire input power range from -30 to 10 dBm and the reflectivity range from -21 to 0 dB, and thus the improved model would allow for accurate and rapid simulations of signal amplification in WDM PON scenarios. The importance of including the internal losses is shown by the red lines in Fig. 4 with the parameters in Table III adjusted to allow for the same net small signal gain that generated the blue curves i.e. $h_0 = 6$ and $\alpha_{loss} = 0$. The results from both the reduced and TWM models show excellent agreement and show a departure from the measured RSOA behavior in [10],[12], thus highlighting the necessity to include internal RSOA losses.

## IV. CONCLUSION

We presented an improved reduced model for both SOAs and RSOAs showing that rapid and accurate simulations can be achieved in a single calculation step.

REFERENCES

[1] G. P. Agrawal and N. A. Olsson, "self-phase modulation and spectral broadening of optical pulses in semiconductor laser amplifiers," *IEEE J. of Quant. Electron.*, 25, no. 11, 2297, (1989).
[2] D. Cassioli, S. Scotti, and A. Mecozzi, "A time-domain computer simulator of the nonlinear response of semiconductor optical amplifiers," *IEEE J. Quantum Electron.*, vol. 36, no. 9, pp. 1072-1080, Sept. 2000.
[3] C. Antonelli and A. Mecozzi, "Reduced Model for the Nonlinear Response of Reflective Semiconductor Optical Amplifiers," *IEEE Photon. Technol. Lett.*, vol. 25, no. 23, pp. 2243 – 2246, Dec 2013.
[4] M. J. Connelly, "Reflective semiconductor optical amplifier pulse propagation model", *IEEE Photon. Technol. Lett.*, vol. 24, no. 2, pp.95 -97, Jan. 2012.
[5] W. R. Lee, M. Y. Park, S. H. Cho, J. Lee, C. Kim, G. Jeong, and B. W. Kim, "Bidirectional WDM-PON based on gain-saturated reflective semiconductor optical amplifiers," *IEEE Photon. Technol. Lett.*, vol. 17, 11, 2460, 2005.
[6] L. Marazzi, P. Parolari, M. Brunero, A. Gatto, M. Martinelli, R. Brenot, S. Barbet, P. Galli, and G. Gavioli, "Up to 10.7-Gb/s High-PDG RSOA-based Colorless Transmitter for WDM Networks," *IEEE Photon. Techn. Lett.*, 25, No. 7, pp. 637-639, 2013.
[7] A. Mecozzi and J. Moerk, "Saturation effects in nondegenerate four-wave mixing between short optical pulses," *IEEE J. of Sel. Topics. In Quant. Electron.*, vol. 3, no. 5, pp 1190 – 1207, Oct. 1997.
[8] R. P. Webb, M. Power, and R. J. Manning, "Phase-sensitive frequency conversion of quadrature modulated signals," *OSA Opt. Expr.*, vol. 21, no. 10, pp. 12713-12727, 2013.
[9] S. T. Naimi, S. Ó Dúill, and L. P. Barry, "Simulations of the OSNR and laser linewidth limits for reliable wavelength conversion of DQPSK signals using four-wave mixing," *Opt. Comms.*, vol. 310, no. 1, pp. 150-155, Jan. 2014.
[10] A. Naughton, C. Antony, P. Ossieur, S. Porto, G. Talli, and P. D. Townsend, "Optimisation of SOA-REAMs for Hybrid DWDM-TDMA PON Applications," *OSA Opt. Expr.*, vol. 19, no. 26, pp. B722-B727, Dec. 2011.
[11] L. W. Casperson, and J. M. Casperson, "Power self-regulation in double-pass high-gain laser amplifiers", *J. Appl. Phys*. vol. 87, no. 5, pp. 2079-2083, 2000.
[12] S. O'Duill, L. Marazzi, P. Parolari, C. Koos, W. Freude and J. Leuthold "Efficient modulation cancellation using reflective SOAs ," *OSA Opt. Expr.*, vol. 20, No. 20, pp. B587-B594, 2012.
[13] G. Eisenstein, G. Raybon, and L. W. Stulz, "Deposition and measurements of electron-beam-evaporated SiOx antireflection coatings on InGaAsP injection laser facets," *IEEE J. of Lightw. Technol.*, vol. 6, no. 1, pp. 12-16, Jan. 1988.